\documentclass[prX,twocolumn,showpacs,showkeys,amsmath,amssymb]{revtex4-2}
\def\oper{{\mathchoice{\rm 1\mskip-4mu l}{\rm 1\mskip-4mu l}
{\rm 1\mskip-4.5mu l}{\rm 1\mskip-5mu l}}}
\def\<{\langle}
\def\>{\rangle}
\usepackage[T1]{fontenc}
\usepackage[utf8]{inputenc}
\usepackage{mathtools,amsthm}
\usepackage{dsfont,bbm}
\usepackage{bm,soul}
\usepackage[scr=boondoxo,scrscaled=1.05]{mathalfa}
\usepackage{dcolumn}% Align table columns on decimal point
\usepackage{graphicx}% Include figure files
\usepackage{float}
\usepackage{subcaption}
\usepackage{xcolor}
\usepackage{enumitem}

\usepackage[normalem]{ulem}

\usepackage{hyperref}
\hypersetup{
	colorlinks   = true, %Colours links instead of ugly boxes
	urlcolor     = darkgreen, %Colour for external hyperlinks
	linkcolor    = darkblue, %Colour of internal links
	citecolor   = darkred, %Colour of citations
	breaklinks = true
}
\colorlet{darkred}{red!55!black}
\colorlet{darkgreen}{green!25!black}
\colorlet{darkblue}{blue!60!black}
\newtheorem*{proposition}{Proposition}
\newtheorem*{remark}{Remark}

\renewcommand{\eprint}[2][]{\href{https://arxiv.org/abs/#2}{arXiv:~\nolinkurl{#2}}}  % arXiv eprint

\begin{document}
	
	%\preprint{}
	
	\title{Universal  bound on the relaxation rates for quantum Markovian dynamics}
	
	\author{Paolo Muratore-Ginanneschi}
	\email{paolo.muratore-ginanneschi@helsinki.fi}
	\affiliation{Department of Mathematics and Statistics, University of Helsinki PL 68, FI-00014, Finland}
 \author{Gen Kimura}
	\email{gen@shibaura-it.ac.jp}
	\affiliation{College of Systems Engineering and Science, Shibaura Institute of Technology, Saitama 330-8570, Japan}
	\author{Dariusz Chru\'sci\'nski}
	\email{darch@fizyka.umk.pl}
	\affiliation{Institute of Physics, Faculty of Physics, Astronomy and Informatics, Nicolaus Copernicus University, Grudziadzka 5/7, 87-100 Toru\'n, Poland}
	\date{\today}% It is always \today, today,
	
	\begin{abstract}
Relaxation rates provide important characteristics both for classical and quantum processes. Essentially they control how fast the system thermalizes, equilibrates, {decoheres, and/or dissipates}. Moreover, very often they are directly accessible to be measured in the laboratory and hence they define key physical properties of the system. Experimentally measured relaxation rates can be used to test validity of a particular theoretical model. 
Here we analyze a fundamental question: {\em does quantum mechanics provide any nontrivial constraint for relaxation rates?} We prove the conjecture formulated a few years ago that any quantum channel implies that a maximal rate is bounded from above by the sum of all the relaxation rates divided by the dimension of the Hilbert space. It should be stressed that this constraint is universal (it is valid for all quantum systems with finite number of energy levels) and it is tight (cannot be improved).  In addition, the  constraint  plays an analogous role to the seminal Bell inequalities and the well known Leggett-Garg inequalities  (sometimes called temporal Bell inequalities). Violations of Bell inequalities rule out local hidden variable models, and violations of Leggett-Garg inequalities rule out macrorealism.
Similarly, violations of the bound rule out completely  positive-divisible evolution.
	\end{abstract}
	

	\maketitle
	
	\section{Introduction}

Time evolution of closed quantum systems is described by unitary maps. %, \gen{based on the Schr\"odinger equation}.
In any realistic scenario, however,  a quantum system is never perfectly isolated, and the interactions between the system and its environment give rise to non-unitary system dynamics causing dissipation, decay, and decoherence. Therefore, the theory of open quantum systems \cite{Davies1976,BrPe2002,RiHu2012} is of primary importance not only for proper description of dynamical features but also for many practical applications to modern quantum technologies and for providing efficient control protocols \cite{WiMi2009}.

The evolution of open quantum systems is represented by quantum channels which transform quantum states (represented by density, or state, operators) according to the laws of quantum physics. Quantum channels (often called quantum maps) are basic objects of quantum information theory \cite{NiCh2010,Wilde2013,Watrous2018}. Mathematically, they are represented by completely positive trace-preserving maps \cite{Kraus1983,Paulsen,LinG1976,GoKoSu1976,Davies1976,Alicki-Lendi}.  
Any such map can be always constructed starting from a unitary map of the ``system + environment'' and then neglecting all degrees of freedom of the environment, the last step being mathematically realized by a partial trace operation. The detailed characteristics of the dynamical properties of open quantum systems are very challenging. The environment usually carries infinite number of degrees of freedom and in the course of evolution system and environment exchange energy and information. The proper description of all these processes is very demanding and without appropriate approximations the problem is rather untractable. 
	
A major breakthrough came in the mid 1970s when Lindblad \cite{LinG1976} and Gorini-Kossakowski-Sudarshan \cite{GoKoSu1976} independently identified the necessary and sufficient conditions for a linear system of differential equations to generate a completely positive semigroup. All particular solutions of  master equations satisfying such conditions, commonly referred to as  Gorini-Kossakowski-Lindblad-Sudarshan {(GKLS)}, or completely positive master equations, are thus quantum channels (cf. \cite{GKLS} for a brief history of this seminal result). 
Traditionally, completely positive semigroups are referred to as  ``Markovian''. The terminology possibly refers to the fact that the master equations of the corresponding form can be derived from microscopic unitary dynamics in the van Hove scaling limit \cite{DavE1974} which rigorously justifies the so called Born-Markov approximation.
Quantum Markovian semigroups describe a plethora of important processes ranging from quantum optical systems \cite{BrPe2002}, control of quantum information processing devices \cite{WiMi2009}, and heat transfer in solid state quantum integrated circuits \cite{PeKa2021}.

A typical measurement provides information about relaxation rates which essentially control the speed of thermalization, equilibration, decoherence, and/or dissipation  \cite{Alicki-Lendi,BrPe2002,RiHu2012}.
Recently, the authors of \cite{ChKiKoSh2021} (see also \cite{KimG2002} for an earlier formulation)  put forward a conjecture concerning a universal relation that the relaxation rates of a quantum process described by completely positive master equation must satisfy. As relaxation rates are experimentally measurable, the conjecture, if generically true, provides an efficient, laboratory realizable protocol to determine whether a large class of quantum channels is specified by a completely positive semigroup.  It provides, therefore, a physical manifestation of the very concept of complete positivity. 
In the present work we prove the conjecture in full generality. Interestingly, the proof draws 
from the theory of Lyapunov exponents, 
a central tool in the analysis of classical dynamical systems \cite{OseV1968,BeGaGiSt1980, SkoC2008,Pikovsky2016}, {especially} in the study of classical \cite{Devaney1986,CoEc2006,CeCeVu2009}, and quantum chaos \cite{Gutzwiller1991,Stoeckmann1999,Haake2018}, and fully developed turbulence theory \cite{BoJePaVu2005}.
In the context of quantum physics, Lyapunov exponents have also been applied to the analysis of quantum channels \cite{Daniel-1}, many body strongly correlated systems \cite{Gharibyan2019}, and  out-of-time order correlators \cite{Bergamasco2023}.  In particular, \cite{Maldacena2016} conjectures the existence of a universal bound on the maximal Lyapunov exponent controlling how fast out-of-time-order correlation functions can grow.

To further underline the general relevance of our result, we recall that not all quantum channels can be expressed as the image of a state operator produced by a completely positive semigroup. This is because the most general form of a quantum master equation stemming from microscopic unitary dynamics does not satisfy the Gorini-Kossakowski-Lindblad-Sudarshan conditions \cite{ShiT1970,ChKo2010,RiHu2012,HaCrLiAn2014}. In such a case, a completely positive evolution may emerge as the image of the ``completely bounded'' semigroup \cite{Paulsen} solving the most general quantum master equation \cite{DoMG2022,DoMG2023,DoMG2023a}.
% In the most general scenario
Evolution processes corresponding to the most general scenario are referred to as ``non-Markovian'' \cite{NM1,NM2,NM3,NM4,ChrD2022}. In this paper by Markovian we mean an evolution which is represented by a dynamical map being a composition of completely positive propagators (very often one calls such evolution CP-divisible \cite{NM1,NM2,NM4,ChrD2022}. For an intricate relation between different concepts of Markovianity cf. \cite{NM4}). These considerations highlight the paramount relevance 
of identifying criteria in terms of experimentally measurable quantities that discriminate between quantum channels generated by Markovian and non-Markovian processes \cite{WoEiCuCi2008}. 
The bound on relaxation rates that we prove enjoys these properties and is independent of the details of the open quantum system dynamics.
Owing to its universality, the result plays an analogous role to Bell inequalities {\cite{Bell2004}} and Leggett-Garg inequalities \cite{LG} (sometimes called temporal Bell inequalities; cf. review article \cite{LG-Nori}): The violation of a Bell inequality rules out local hidden variable models, and the violation of Leggett-Garg inequalities rules out classical descriptions. Similarly, the violation of the relaxation bound rules out completely positive Markovian dynamics.

The structure of the paper is as follows. In section~\ref{sec:conjecture} we explain the mathematical contents of the conjecture.
We recall the results known so far which substantiate the conjecture.  Section~\ref{sec:proof} expounds the main result of the present work. After some mathematical preliminaries we show that the proof in full generality of the conjecture follows from the Teich-Mahler \cite{TeMa1992} construction
of the Pauli master equation equivalent to a canonical quantum master equation. Such construction has the advantage to decouple from the time asymptotic behavior of the solution degrees of freedom associated to rotations of the eigenvectors of the state
operators. Hence, looking at the dynamics for large negative times allows us to derive an optimal bound on the relaxation rates by rephrasing the problem in terms of Lyapunov exponents. Tightness of the bound stems from results of \cite{ChKiKoSh2021}. It is worth emphasizing here that in general finding tight bounds for maximal Lyapunov exponents and especially for cocycles is a mathematically very hard problem \cite{TsBl1997}. It is a surprising and non-trivial consequence of the structure of the Lindblad generators that allows us to find a tight bound.
 In section~\ref{sec:nM} {we observe that the bound still holds true for time dependent generators giving rise to Markovian evolution corresponding to completely positive divisible dynamical maps. Hence, violation of the bound provides a clear witness of non-Markovianity.} Finally Section~\ref{sec:outlook} provides further discussion directly related to the original bound. In particular in  section~\ref{sec:Lyapunov} we turn the attention to time non-autonomous quantum master equations. There we show that rephrasing the conjecture in terms of Lyapunov exponents produces a bound that distinguishes the qualitative properties of completely positive divisible master (Markovian) equations from those of non-Markovian ones. In particular, we recover the role of the trace-norm rate of change introduced in \cite{RiHuPl2010} as a witness of non-completely positive evolution.
Section~\ref{sec:finale} is devoted to the conclusions while further technical information is deferred to the appendices.

\section{The Conjecture and its implications}
\label{sec:conjecture}

We consider the canonical (diagonal) form of the time-autonomous {GKLS} master equation for a density operator ${\bm{\rho}(t)}$ of an open quantum system with $d$-dimensional Hilbert space, that is, $${\dot{\bm{\rho}}(t)= \mathfrak{L}(\bm{\rho}(t)),}$$ where 
	\begin{align}
	\mathfrak{L}(\bm{{\rho}}) =-\imath\,\left[\operatorname{H},\bm{\rho}\right]+
\sum_{\ell=1}^{d^{2}-1}\gamma_{\ell} \left( \operatorname{L}_{\ell}\bm{\rho} \operatorname{L}_{\ell}^\dagger - \frac 12 \{  \operatorname{L}_{\ell}^\dagger \operatorname{L}_{\ell},\bm{\rho}\} \right) ,
		\label{LGKS}
	\end{align}
{Here, $\operatorname{H}$ is a Hermitian operator describing an effective Hamiltonian of the system. To ensure complete positivity, the canonical transition rates $\gamma_{\ell}$ must satisfy the positivity condition}:
\begin{align}
\gamma_{\ell} \,\geq\,0 \ , \ \ \ \ell=1,\dots,d^{2}-1. 
		\label{cp}
\end{align}
{The noise (Lindblad)} operators $\operatorname{L}_{\ell}$ control the coupling between the system and the environment. {In this paper, the imaginary unit is denoted by the symbol `$\imath$'. 
The GKLS generator gives rise to completely positive trace-preserving dynamical map $\{\Lambda(t)  = \exp(t \mathfrak{L}) \}_{t \geq 0}$ which describes the evolution in the Schr\"odinger picture: $$\bm{\rho}(0) \to \bm{\rho}(t) = \Lambda(t)(\bm{\rho}(0)).$$ A stationary state $\bm{\rho}_{\rm SS}$ of the corresponding evolution satisfies $\mathfrak{L}(\bm{\rho}_{\rm SS})=0$, that is, it defines a zero-mode of $\mathfrak{L}$.
Denote by $\lambda_\ell$ ($\ell=1,\ldots,d^2-1$) the remaining eigenvalues of $\mathfrak{L}$: 
$$
\mathfrak{L}(X_\ell) = \lambda_\ell X_\ell.
$$
It is well known \cite{ChrD2022,WolM2012} that eigenvalues are in general complex, however, the spectrum is symmetric w.r.t. real line, if $\lambda_\ell$ belongs to the spectrum so does $\lambda_\ell^*$. The key property of the spectrum is that %${\rm Re}\lambda_\ell \leq 0$, i.e. 
all eigenvalues are located on the left half of the complex plane. In other words, the eigenvalues are real negative or appear in complex conjugated pairs with negative real part:
	\begin{align}
	&	\Gamma_{\ell}=-\operatorname{Re}\lambda_{\ell}\,\geq\,0&&\ell =0,\dots,d^{2}-1
		\label{rates}
	\end{align}
One calls $  \Gamma_{\ell}$'s  the relaxation rates. Equivalently $$T_\ell = 1/\Gamma_\ell $$ are called relaxation times.
 We label the eigenvalues in increasing order of the magnitude of the relaxation rates. We count them starting from zero to emphasize that the spectrum always includes (see section~2.5 of \cite{ChrD2022})  at least one vanishing eigenvalue
    \begin{align}
    	0=\lambda_{0}=\Gamma_{0}\,\leq\,\Gamma_{1}\,\leq\,\dots\,\leq\,\Gamma_{d^{2}-1} .
    	\nonumber
    \end{align}
The \emph{universal bound} \cite{ChKiKoSh2021}, which we set out to prove is  
\begin{align}
    	\Gamma_{d^{2}-1}\,\leq\,\frac{1}{d}\sum_{\ell=1}^{d^{2}-1}\Gamma_{\ell} ,
    	\label{bound}
    \end{align}
that is, the maximal relaxation rate is always upper-bounded by the total rate $\Gamma := \sum_\ell \Gamma_\ell$ divided by the dimension of the system's Hilbert space. 
{{The bound (\ref{bound}) provides a multi-level generalization of a seminal relation between longitudinal and transversal rates for qubit evolution governed by the following Lindblad generator 
\begin{equation}
\label{QUBIT-1}
\begin{split}
&  \mathfrak{L}(\bm{\rho}) = - \imath\,\frac{\omega}{2}[\sigma_z,\bm{\rho}] +\sum_{i=+,-,z}\gamma_i \mathfrak{D}[\sigma_{i}](\bm{\rho}), 
  \\
 & \mathfrak{D}[\sigma_{i}](\bm{\rho}) = \sigma_{i}\bm{\rho}\sigma_{i}^{\dagger}
- \frac 12 \left( \sigma_{i}^{\dagger}\sigma_{i}\bm{\rho}+\bm{\rho}\sigma_{i}^{\dagger}\sigma_{i} \right) .
\end{split}
\end{equation}
As usual $$\sigma_\pm = (\sigma_x \pm \imath \sigma_y)/2$$ are raising/lowering qubit operators and $\sigma_{x}$, $\sigma_{y}$, and $\sigma_{z}$ are the Pauli matrices.  The corresponding relaxation rates read
\begin{align}%\label{QUBIT:rates}
&\Gamma_{\rm L} =  \gamma_{+} + \gamma_{-}  & (\mbox{longitudinal})
\nonumber  \\
 & \Gamma_{\rm T} = \frac{\gamma_{+} + \gamma_{-}}{2} + \gamma_{z}   &(\mbox{transversal, 2-degenerate}) .
\nonumber
\end{align}
The total rate equals to $\Gamma = \Gamma_{\rm L} + 2 \,\Gamma_{\rm T} $ and hence (\ref{bound}) reduces the the following well known condition $2 \Gamma_{\rm T} \geq \Gamma_{\rm L}$, or, equivalently, in terms of local relaxation times {\cite{Abragam1961,Slichter1990,GoKoSu1976,Alicki-Lendi}}
\begin{equation}\label{TT-1}
% 2\, T_{{\rm L}}(t) \geq  T_{\rm T}(t) .
 2\, T_{\rm L}\geq  T_{\rm T} .
\end{equation}
This result clearly shows that complete positivity implies nontrivial constraint between relaxation rates (or equivalently relaxation times). Note that introducing the corresponding Bloch vector
\begin{equation}
 \bm{r} = (x,y,z) = \left( {\rm Tr}(\sigma_x \bm{\rho}), {\rm Tr}(\sigma_y \bm{\rho}),{\rm Tr}(\sigma_z \bm{\rho}) \right) ,
\end{equation}
one finds that Lindblad master equation (\ref{QUBIT-1}) is equivalent to the following Bloch equations
\begin{equation}  
\label{Bloch-eqs}
\begin{bmatrix}
\dot{x}(t)	\\ \dot{y}(t) \\ \dot{z}(t) 
\end{bmatrix}
=
\begin{bmatrix}
	- \Gamma_{\rm T} & - \omega & 0 \\   \omega & - \Gamma_{\rm T} & 0 \\ 0 & 0 &  - \Gamma_{\rm L}
\end{bmatrix}
\begin{bmatrix}
x(t)	\\ y(t) \\ z(t) 
\end{bmatrix}
+\begin{bmatrix}
0 \\ 0 \\  \delta
\end{bmatrix}
\end{equation}
with $\delta = \gamma_+ - \gamma_- $. If $t \to \infty$ one finds $\bm{r}(t)\to (0,0,\delta/\Gamma_{\rm L})$. 
Now, if $\bm{r}(0)$ belongs to the Bloch ball (i.e. $\|\bm{r}(0)\|_{2}\leq 1$, where $\|\cdot\|_{2}$ denotes the Euclidean norm), then $\bm{r}(0)$ stays in the Bloch ball whenever $\Gamma_{\rm T} \geq 0$ and $\Gamma_{\rm L} \geq |\delta|$. Hence, the very condition (\ref{TT}) is hidden in the standard Bloch representation. Note, however, that Bloch representation does not care about complete positivity. If in the course of time $\|\bm{r}(t)\|_{2} \leq 1$, then evolution of $\bm{\rho}(t)$ is necessarily positive. However, complete positivity is not guaranteed. Necessary and sufficient condition for complete positivity reads
\begin{equation}
    2\, \Gamma_{\rm T} \geq \Gamma_{\rm L} \geq |\delta| ,
\end{equation}
which immediately implies the very constraint (\ref{TT-1}).  Note, that the parameter $\delta$ controls how much the evolution breaks the unitality. For the unital case $\delta=0$ condition (\ref{TT-1}) is therefore necessary and sufficient for complete positivity. This simple example nicely illustrates how the very requirement of complete positivity gives rise to highly nontrivial additional constraints for relaxation rates.}}

A characteristic trait of the canonical form (\ref{LGKS}) is that the decoherence operators form an orthonormal basis with respect to the Hilbert-Schmidt inner product of the restriction of $\mathcal{M}_{d}(\mathbb{C})$ (the space of $d\,\times\,d$ complex matrices) to traceless elements \cite{HaCrLiAn2014}
	\begin{equation}
		\begin{split}
		&\operatorname{Tr}\operatorname{L}_{\ell}^{\dagger}\operatorname{L}_{\mathscr{k}}=\delta_{\ell,\mathscr{k}}
		\\
		&\operatorname{Tr}\operatorname{L}_{\ell}=0					
		\end{split}
		\hspace{1.0cm}\forall\, \ell,\mathscr{k}=1,\dots,d^{2}-1
		\label{ortho}
	\end{equation}
In this case one has $$\frac{1}{d}\sum_{\ell=1}^{d^{2}-1}\Gamma_{\ell} = \sum_{\ell=1}^{d^{2}-1}\gamma_{\ell}$$ (cf. \cite{WoCi2008,KiAiWa2017,ChKiKoSh2021}). In \cite{KimG2002} one of us proved the bound for the qubit completely positive master equation.
    In \cite{ChKiKoSh2021} the authors proved (\ref{bound}) under additional hypotheses, most prominently when the evolution is unital, i.e. $$\mathfrak{L}(\oper) = 0.$$ Such quantum evolution preserves maximally mixed state. In terms of Lindblad noise operators $\operatorname{L}_n$ it corresponds to $$\sum_{n=1}^{d^{2}-1}\gamma_{n}\,[\operatorname{L}_{n}^{\dagger}\,,\operatorname{L}_{n}]=0 .$$ Interestingly, the bound was also proved when the generator displays the following covariance property \cite{ChKiKoSh2021}
\begin{align}
    \operatorname{U}\mathfrak{L}(\bm{\rho})\operatorname{U}^\dagger = \mathfrak{L}(\operatorname{U}\bm{\rho}\operatorname{U}^\dagger) ,
\end{align}
for all $\operatorname{U}$ belonging to the maximal commutative subgroup of the full unitary group $U(d)$. Finally, the bound was proved for a class of generators derived in the weak coupling limit from the proper microscopic model describing the system-environment interaction -- so-called Davies generators \cite{DavE1974,Davies-2,Davies-3,Davies1976}. Such class of generators provides basic tool to analyze majority of quantum optical systems \cite{BrPe2002}.  Furthermore, \cite{ChKiKoSh2021} gives an example when (\ref{bound}) holds as an equality hence implying tightness. Non-tight general upper bounds are otherwise implied by results in \cite{WoCi2008}, \cite{KiAiWa2017},  and \cite{ChFuKiOh2021}. We list these references in increasing order of refinement of the bound.

{{
\subsection{Classical vs. quantum semigroups}
A classical counterpart of the Markovian master equation is provided by the well known classical master equation for the probability vector $\bm{p}(t) {\in \mathbb{R}^d}$
\begin{equation}   \label{CL}
    \dot{\bm{p}}(t) = \operatorname{K} \bm{p}(t) ,
\end{equation}
where $\operatorname{K} {= (\operatorname{K}_{ij})}$ is a real $d \times d$ matrix satisfying the following conditions \cite{Kampen2007}
\begin{equation}  \label{K1}
    \operatorname{K}_{ij} \geq 0 \ , \ \ (i \neq j) \ ,
\end{equation}
and
\begin{equation}  \label{K2}
    \sum_{i=1}^d \operatorname{K}_{ij} = 0 . 
\end{equation}
Any Kolmogorov generator {$\operatorname{K}$} can be represented as follows
\begin{equation}
    \operatorname{K}_{ij} = \operatorname{R}_{ij} - \delta_{ij} \sum_{k=1}^d \operatorname{R}_{kj} ,
\end{equation}
with nonnegative rates $\operatorname{R}_{ij}$ (note that diagonal elements $\operatorname{R}_{ii}$ do not contribute to $\operatorname{K}_{ij}$). This way (\ref{CL}) can be equivalently rewritten in the form of the Pauli rate equation
\begin{equation}   \label{Pauli-0}
    \dot{p}_i(t) = \sum_{j=1}^d \Big( \operatorname{R}_{ij} p_j(t) - \operatorname{R}_{ji} p_i(t) \Big) .
\end{equation}
Interestingly, given a Lindblad generator $\mathfrak{L}$ and fixing an arbitrary orthonormal basis in the system's Hilbert space $\{|1\>,\ldots,|d\>\}$ the following matrix
\begin{equation}
    \operatorname{K}_{ij} := \<i| \mathfrak{L}(|j\>\<j|)|i\> , 
\end{equation}
satisfies (\ref{K1}) and (\ref{K2}). Moreover, the transition rate matrix $ \operatorname{R}_{ij}$ reads as follows
\begin{equation}   \label{Rij-0}
    \operatorname{R}_{ij} := \sum_{n} \gamma_n |\< i| \operatorname{L}_n|j\>|^2 .
\end{equation}
The spectrum of a classical generator $\operatorname{K}$ has similar properties as the spectrum of a quantum Lindbladian: eigenvalues $\{\ell_i\}_{i=0}^{d-1}$ are either real or appear in complex conjugate pairs. Moreover $\ell_0=0$ and the remaining eigenvalues satisfy ${\rm Re}\, \ell_i\leq 0$. One defines classical relaxation rates $r_i := - {\rm Re}\, \ell_i$. Is there any nontrivial constraint among classical relaxation rates? Interestingly, contrary to the quantum case governed by the Lindblad generator $\mathfrak{L}$ one has the following

\begin{proposition}\cite{ChKiKoSh2021} Given an arbitrary set of nonnegative numbers $\{r_1,\ldots,r_{d-1}\}$ there exists a classical generator $\operatorname{K}_{ij}$ such that $\{r_i\}_{i=1}^{d-1}$ define its relaxation rates.     
\end{proposition}
\emph{Proof}: {To} prove this result let us construct the corresponding $K_{ij}$ 
\begin{equation} \label{K!}
    \operatorname{K} =
\left(
\begin{array}{cccccc}
-r_1 & r_2 & r_3 & \cdots & r_{d-1} & 0  \\
0 & - r_2 & 0 & \cdots & 0 & 0
\\
0 & 0 & -r_3 & \cdots& 0  & 0
\\
\vdots & \cdots & \vdots & \ddots & \vdots & \vdots \\
0 & 0 & 0 & \cdots& -r_{d-1}  & 0
\\
r_1 & 0 & \cdots & 0 & 0   & 0
\end{array}
\right) ,
\end{equation}
which obviously satisfies (\ref{K1}) and (\ref{K2}) and its spectrum $\sigma(\operatorname{K}) =\{0,-r_1,\ldots,-r_k\}$, that is, numbers  $\{r_1,\ldots,r_{d-1}\}$ define relaxation rates of (\ref{K!}). Note that (\ref{K!}) does generate very particular classical process in which state `$i$' ($i=2,\ldots,d-1$) jumps to the state `$1$' with the rate $r_i$, and the state `$1$' jumps to `$d$' with the rate $r_1$. It should be stressed that there are other generators with relaxation rates $r_i$ and (\ref{K!}) is just an example.   
 }}

\section{Proof of the conjecture}
\label{sec:proof}

\subsection{Mathematical preliminaries}

The dynamics of state operators (\ref{LGKS}) is fully specified by the
solution of	the spectral problem
	\begin{align}
		&		\begin{split}
			\mathfrak{L}(\operatorname{X}_{\ell})=\lambda_{\ell}\,\operatorname{X}_{\ell}
			\\
			\mathfrak{L}^{\ddagger}(Y_{\ell}^{\dagger})=\bar{\lambda}_{\ell}\,Y_{\ell}^{\dagger}
		\end{split}
		&& \ell=0,\dots,d^{2}-1 ,
		\label{sp}
	\end{align}
where $ \mathfrak{L}^{\ddagger}$ is the adjoint of $\mathfrak{L} $ with respect to the Hilbert-Schmidt inner product on $\mathcal{M}_{d}(\mathbb{C})$, that is, $${\rm Tr}(X \mathfrak{L}(Y)) = {\rm Tr}(\mathfrak{L}^{\ddagger}(X) Y)$$ for all $X,Y \in \mathcal{M}_{d}(\mathbb{C})$. Note, that
\begin{align}
	\mathfrak{L}^\ddagger(\bm{X}) =\imath\,\left[\operatorname{H},\bm{X}\right] + 
\sum_{\ell=1}^{d^{2}-1}\gamma_{\ell} \left( \operatorname{L}_{\ell}^\dagger\bm{\rho} \operatorname{L}_{\ell} - \frac 12 \{  \operatorname{L}_{\ell}^\dagger \operatorname{L}_{\ell},\bm{\rho}\} \right) ,
		\nonumber
\end{align}
generates the evolution in the Heisenberg picture $$\Lambda^{\ddagger} (t)= \exp(t \mathfrak{L}^\ddagger)$$ such that
\begin{align}
 {\rm Tr}\Big( \bm{X} \Lambda(t)(\bm{\rho})\Big) = {\rm Tr}\Big(\Lambda^{\ddagger}(t)({\bm X}) \bm{\rho} \Big)
\end{align}
for any state operator $\bm{\rho}$ and any observable $\bm{X}$. The above formula essentially states the equivalence between the Schr\"odinger and Heisenberg  pictures beyond the standard unitary scenario.

Consider now the generic case when  $ \mathfrak{L}$ is diagonalizable (non defective), i.e. right eigenvectors $\operatorname{X}_\ell$ (equivalently, left eigenvectors $Y_\ell$) are linearly independent. Since diagonalizable generators are dense in the space of all generators, {proving the conjecture requires only considering a generic case. (In Appendix~\ref{app:spectrum}, we demonstrate how to handle defective generators).} Interestingly, generators {that} are not diagonalizable (defective generators) recently raised {considerable} attention due {to} the the presence of {so-called} exceptional points \cite{,Miranowicz2019,Miranowicz-2023}. {The latter correspond to eigenvalues $\lambda_\ell$ associated with} nontrivial Jordan blocks (cf. Appendix~\ref{app:spectrum}).  Any diagonalizable generator has the following spectral representation

 \begin{align}
     \mathfrak{L}(\bm{\rho}) = \sum_{\ell=0}^{d^2-1} \lambda_{\ell} \operatorname{X}_{\ell}\operatorname{Tr}
	\left(Y_{\ell}^{\dagger} \bm{\rho}\right) .
 \end{align}
Hence,  we are entitled to generically write the spectral representation of the corresponding dynamical map $\Lambda(t) = e^{t \mathfrak{L}}$ as follows
\begin{align}
	\bm{\rho}(t) = \Lambda(t)({\bm{\rho}(0)})=  \sum_{\ell=0}^{d^{2}-1}e^{\lambda_{\ell}\,t}\operatorname{X}_{\ell}\operatorname{Tr}
	\left(Y_{\ell}^{\dagger} \bm{\rho}(0) \right) .
	\label{stop}
\end{align}  	

{
\begin{remark}
In was shown \cite{ChKiKoSh2021} that for diagonalizable generators the relaxation rates $\Gamma_\ell$ satisfy the following identity
\begin{equation} 
    \Gamma_\ell = \frac{1}{2\| Y_\ell \|_{\rm ss}^2 } \sum_{k=1}^{d^2-1} \gamma_k \| [\operatorname{L}_k,Y_\ell] \|^2_{\rm ss} ,
\end{equation}
where 
$$  \| Y \|_{\rm ss}^2 :=  {\rm Tr}( \bm{\rho}_{\rm ss} Y^\dagger Y) , $$
and $\bm{\rho}_{\rm ss}$ stands for the stationary state. Now, if $\bm{\rho}_{\rm ss}$ is maximally mixed (i.e. the corresponding semigroup is unital), then  $\| Y \|_{\rm ss}^2 = \frac 1d {\rm Tr}(Y^\dagger Y) = \frac 1d \|Y\|^2_2$ reduces to the standard Hilbert-Schmidt norm and hence using the B\"otcher-Wenzel inequality \cite{BW}
\begin{equation}   \label{BW}
    \| [A,B]\|^2_2 \leq 2\|A\|^2_2 \|B\|^2_2 ,
\end{equation}
together with $\|\operatorname{L}_\ell\|_2 =1$, one arrives at (\ref{bound}). Unfortunately, the above proof no longer works beyond the unital scenario. Possible generalization of B\"otcher-Wenzel inequality were recently analyzed in \cite{Aina2024}.
\end{remark}
}

 The semi-group generated by (\ref{LGKS}) by considering forward in-time evolution is completely positive. This property is lost if we instead consider the one-parameter family group of transformations (flow) obtained by including backward in time evolution (see e.g. Theorem~3.4.1 of \cite{RiHu2012}). Nevertheless, as (\ref{LGKS}) is equivalent to a system of differential equations on ${\mathcal{M}_d(\mathbb{C}) \simeq} \mathbb{C}^{d^{2}}$ the flow of (\ref{LGKS}) is well-defined for all times. In addition, the flow always preserves the trace and self-adjoint property of the initial conditions. {As we will see, the key to proving the bound \eqref{bound} lies in tracking the negative times.} 

    \subsection{Mapping onto an effective Pauli master equation}
    Any snapshot of a self-adjoint solution of (\ref{LGKS}) admits the diagonal representation
      \begin{align}
     	\bm{\rho}(t) = \sum_{i=1}^{d}\wp_i(t) |\bm{\psi}_i(t)\> \<\bm{\psi}_i(t)| .
     	\label{diagonal}
     \end{align}
     The collection $\big{\{}\bm{\psi}_i(t)\big{\}}_{i=1}^{d}$  specifies a complete orthonormal basis of $\mathbb{C}^{d}$ at any instant of time $t$. Hence, the eigenvalues $\big{\{}\wp_i(t) \big{\}}_{i=1}^{d}$ can be thought as the entries of a $d$-dimensional} vector $\bm{\wp}_t\in \mathbb{R}^{d}$, satisfying the trace preservation condition
     \begin{align}
     	\sum_{i=1}^{d}\wp_i(t)=1 ,
     	\label{trace}
     \end{align}
     provided they satisfy it at $t=0$. For our argument, it is not restrictive to assume that this is always the case. By taking advantage of the fact that complete orthonormal bases of a vector space are connected by a unitary transformation, Teich and Mahler showed in \cite{TeMa1992} (see also \cite{WiTo1999}) that the $\bm{\wp}_{t}$ satisfies the  linear time-non-autonomous system of equations
     \begin{align}
     	\bm{\dot{\wp}}(t) = \operatorname{W}(t)\bm{\wp}(t) ,
     	\label{TM}
     \end{align}
     specified by $d \times d$ real matrix $\operatorname{W}(t)$ with entries
     \begin{align}
     	\operatorname{W}_{ij}(t) = \operatorname{R}_{ij}(t) - \delta_{ij}\, \sum_{k=1}^d  \operatorname{R}_{kj}(t) ,
     	\label{classical}
     \end{align}
where we use the definition
\begin{align}
    \operatorname{R}_{ij}(t)= \sum\limits_{n=1}^{d^{2}-1}\gamma_{n}  \left| \left\langle\,\bm{\psi}_i(t) \,,\operatorname{L}_{n}\bm{\psi}_j(t)\,\right\rangle %_{d}
    \right |^{2} .
\end{align}
Note that $R_{ij}(t)$ is just time-dependent version of (\ref{Rij-0}). The complete positivity condition \eqref{cp} and the above definition immediately imply that for each $t$ one has $$\operatorname{R}_{ij}(t) \geq 0, $$ and $$\sum_{i=1}^d \operatorname{R}_{ij}(t)=0.$$
Actually, complete positivity is sufficient but not necessary to have the above properties. Interestingly, positivity is sufficient. Hence (\ref{TM}) defines a classical Pauli rate equation for a probability vector
\begin{align}  \label{Pauli-t}
    \dot{\wp}_i(t) = \sum_{j=1}^d \Big( \operatorname{R}_{ij}(t) {\wp}_j(t)  - \operatorname{R}_{ji}(t){\wp}_i(t) \Big) ,
\end{align}
with time-dependent rates $\operatorname{R}_{ij}(t)$. 

The  dynamics of the {$d$}-vector $\bm{\wp}_{t}$ is completely specified by the spectral properties of (\ref{LGKS}). Namely, if we contrast (\ref{diagonal}) with (\ref{stop}), we generically get
     \begin{align}    
     	\wp_{i}(t) = \sum_{\ell=0}^{d^{2}-1} e^{\lambda_{\ell}\,t} \left \langle\,\bm{\psi}_i(t)\,,\operatorname{X}_{\ell}\bm{\psi}_i(t)\,\right\rangle \operatorname{Tr}\left(Y_{\ell}^{\dagger}\bm{\rho}(0)\right) .
     	\label{growth}
     \end{align}

{  
\begin{remark}   Again (\ref{Pauli-t}) is a time dependent analog of (\ref{Pauli-0}). An interesting property of (\ref{Pauli-t}) is that all rates $\operatorname{R}_{ij}(t)$ --though time dependent -- are non negative. 
Using (\ref{growth}) one easily finds the classical dynamical map
\begin{equation}
    \wp(0) \to \wp(t) = \mathscr{F}(t,0) \wp(0) ,
\end{equation}
that is,
\begin{equation}
    \mathscr{F}_{ik}(t,0) = \sum_{\ell=0}^{d^{2}-1} e^{\lambda_{\ell}\,t} \left \langle\,\bm{\psi}_i(t)\,,\operatorname{X}_{\ell}\bm{\psi}_i(t)\,\right\rangle \, \langle \,\bm{\psi}_k(0)\,,Y_{\ell}^{\dagger} \bm{\psi}_k(0)\,\rangle
\end{equation}
defines a family of $d \times d$ stochastic matrices. The map $\mathscr{F}_{t,0} $ has the following divisibility property
\begin{equation}
    \mathscr{F}(t,0) = \mathscr{F}(t,s) \mathscr{F}(s,0) ,
\end{equation}
and the classical propagator $\mathscr{F}(t,s)$ defines a stochastic matrix for any $t \geq s$. If the rates do not depend on time (like in (\ref{Pauli-0}), then the propagator is time homogeneous, that is, $$\mathscr{F}(t,s) = \mathscr{F}(t-s,0) = e^{(t-s) \operatorname{K}}.$$. Hence, quantum time homogeneous dynamics of the density operator $\bm{\rho}(t)$ gives rise to time inhomogeneous but divisible classical evolution of the probability vector $\wp_t$ consisting of eigenvalues of $\bm{\rho}(t)$.    
\end{remark}
}

\subsection{Derivation of the bound via the theory of Lyapunov exponents}

For positive time the Pauli rate equation (\ref{TM})  preserves legitimate probability vectors $\bm{{\wp}}(t)$: (\ref{trace}) holds as the sum of positive elements.  This is no longer true for $t < 0$ when $\bm{{\wp}}(t)$ components may turn negative.
In the large negative time, the above expression is exponentially dominated by the eigenvalue with largest rate
     \begin{align}
%     	\begin{split}
&     	\wp_{i}(t)\overset{t \downarrow -\infty}{\to} e^{-\Gamma_{d^{2}-1}t}\mathscr{q}_{i}(t)%\mathscr{q}_{t}^{(i,d^{2}-1)}
     	+\dots
\label{asy}
     \end{align}
    The dots stand for exponentially smaller corrections vanishing with rate determined by the difference between $ \Gamma_{d^{2}-1}$ and the second largest rate.     The $ \mathscr{q}_i(t)$'s are always real quantities that specify in the generic case bounded functions of time.
   {If $\lambda_{d^2-1} = -i \omega_{d^2-1} - \Gamma_{d^2-1}$ is a simple eigenvalue, then
\begin{equation}   \label{qt}
    \mathscr{q}_i(t) = \left\langle\,\bm{\psi}_i(t)\,, \bm{Q}(t) \bm{\psi}_i(t)\,\right\rangle \,  ,
\end{equation}
with time-dependent Hermitian operator
$$ \bm{Q}(t) = e^{-i \omega_{d^2-1} t}\, y \operatorname{X}_{d^2-1} + e^{i \omega_{d^2-1} t} \, y^*\operatorname{X}^\dagger_{d^2-1} ,$$
and a complex parameter
$$ y = \operatorname{Tr}\left(Y_{d^2-1}^{\dagger}\bm{\rho}(0)\right) ,$$
and it clearly shows that $ \mathscr{q}_i(t)$ is a bounded function of time. }  
In the non-generic defective case (i.e. when $\mathfrak{L}$ is not diagonalizable),  $\mathscr{q}_i(t)$ is at most polynomially increasing. Under any circumstances, trace preservation requires
     \begin{align}
     \sum_{i=1}^{d}\mathscr{q}_{i}(t)=0 ,
     	%\label{ncp}
      \nonumber
     \end{align}
{which is clearly seen e.g. from (\ref{qt}) due to the fact that ${\rm Tr}\operatorname{X}_{\ell} =0$ for $\ell =1,\ldots,d^2-1$. }

Inspection of (\ref{growth}) evidences that finding $\Gamma_{d^{2}-1}$ is equivalent to computing:
     \begin{align}
    \chi := \lim_{t\downarrow -\infty}\sup_{t} \left(-\,\frac{1}{t}\ln \|\bm{\wp}(t)\|\right ),
     	\label{Lyapunov}
     \end{align}
This quantity corresponds to the maximal Lyapunov exponent of the flow {(backward in time)} of (\ref{TM}). 
{Interestingly, it turned out that the maximal relaxation rate $\Gamma_{d^2-1}$ for the quantum master equation defines the maximal Lyapunov exponent for the corresponding classical evolution governed by the time-dependent Pauli rate equation (\ref{TM}), i.e. each ``classical'' trajectory $\bm{{\wp}}(t) $ is governed by the same characteristic exponent $ \chi =\Gamma_{d^{2}-1}$.
}
Modern literature on Lyapunov exponents (see e.g. \cite{Pikovsky2016,CeCeVu2009}) usually refers to Oseledets's non-commutative ergodic theorem \cite{OseV1968}. For our purposes, however, the theory antedating Oseledets' celebrated result,  developed in \cite{ByViGrNe1966} and summarized in \cite{AdrL1995} has a more direct bearing on our case. The main takeaway is that the maximal Lyapunov exponent is an affine invariant of the  flow $\mathscr{F}$ solving (\ref{TM}):
$$ \bm{\wp}(t) = \mathscr{F}(t,0) \bm{\wp}(0)$$
In other words, the limit in (\ref{Lyapunov}) is independent of the choice of the norm provided this latter is related by equivalence inequalities to the Euclidean one. For any matrix norm induced by an equivalent Euclidean vector norm, the flow satisfies the bound
$$
\|\mathscr{F}(t,0)\| \leq \exp \left( \int_{0}^{t} \mathrm{d}t \| \operatorname{W}(t)\| \right) ,
$$
for any real value of $t$ \cite{AdrL1995}. Correspondingly, if the inequality
 $$\sup_t \| \operatorname{W}(t) \| \leq K , $$
 holds true then it certainly implies
 $$ \chi=\Gamma_{d^2-1} \leq K  . $$
Thus, our task ultimately reduces to determining the norm that provides the smallest upper bound on $\chi$.

\subsection{Different norms result in  different bounds}

 %\emph{Different norms result in  different bounds.}
The last step of the proof consists in choosing an appropriate norm. In $\mathbb{R}^d$ there are three natural norms:

\begin{align}
    \|\bm{x}\|_1 = \sum_{i=1}^d |x_i| \ , \ \   \ \  \|\bm{x}\|_\infty = \max_i |x_i| .   \nonumber
\end{align}
and the standard Euclidean norm $ \|\bm{x}\|_2^2 = \sum_{i=1}^d x_i^2$.
Each norm induces the corresponding norm in the space of $d \times d$ real matrices $\operatorname{A}=(a_{ij})$:
\begin{align}
    \| \operatorname{A}\|_1 = \max_j \sum_{i=1}^d |a_{ij}| \ , \ \  \ \ \| \operatorname{A}\|_\infty = \max_i \sum_{j=1}^d |a_{ij}| ,
      \nonumber
\end{align}
and the Euclidean norm induces the spectral norm  $$\|\operatorname{A}\|_2 = \sup_{\|\bm{x}\|_2=1} \| \operatorname{A} \bm{x}\|_2.$$
Let us compute $\|\operatorname{W}(t)\|_\infty$. {Now, assuming that all $\gamma_\ell \geq 0$ (complete positivity)} we find
\begin{align}
&	\|\operatorname{W}(t)\|_\infty = \max_{1\,\leq\,i\,\leq\,d}\sum_{j=1}^{d}|\operatorname{W}_{ij}(t)|\,\leq\,\sum\limits_{n=1}^{d^{2}-1}\gamma_{n}\max_{1\,\leq\,i\,\leq\,d} w_{n}^{(i)}(t) ,
\nonumber
\end{align}
where in the rightmost expression we introduce the following time-dependent quantity
\begin{align}
&w_{n}^{(i)}(t) =
\nonumber\\
&\sum_{j\neq i}^{d}\left(\left |\left \langle\bm{\psi}_j(t)\,,\operatorname{L}_{n}\bm{\psi}_i(t)\right\rangle%_{d}
\right |^{2}+\left |\left \langle\bm{\psi}_j(t)\,,\operatorname{L}_{n}^{\dagger}\bm{\psi}_i(t)\right\rangle %_{d}
\right |^{2}\right)
	\label{X} .
\end{align}
Simple analysis (see Appendix~\ref{app:bound}) shows that   $w_{n}^{(i)}(t)\leq 1$, and hence  $$\|\operatorname{W}(t)\|_\infty \leq \sum\limits_{n=1}^{d^{2}-1}\gamma_{n}$$
which finally implies
\begin{align}
    \Gamma_{d^2-1} = \chi \leq \sum\limits_{n=1}^{d^{2}-1}\gamma_{n} = \frac {1}{d} \sum\limits_{n=1}^{d^{2}-1}\Gamma_{n} ,
\end{align}
and hence  proves the original bound (\ref{bound}).
Since we already know that the bound is tight \cite{ChKiKoSh2021}, any other norm gives rise to a looser bound.
See Appendix~\ref{app:bound} and \ref{app:DL} for further details.

\subsection{Comments to the proof}

Two comments are in order:

\begin{itemize}
    \item The matrix norms $ \| \|_{1}$ and $ \| \|_{\infty}$ are usually referred to as the ``column-sum'' and ``row-sum'', respectively. Gershgorin's circle theorem \cite{VarR2004} allows us to interpret these norms as estimates of the absolute value of real part of largest eigenvalue of the rate matrix $\operatorname{W}(t)$. The optimal bound stems from the fact that for a rate matrix the row-sum norm provides the optimal Gershgorin's circle estimate.

\item Numerical algorithms to compute Lyapunov exponents often rely on the QR representation of the linear flow matrix \cite{SkoC2008,Pikovsky2016}. In Appendix~\ref{app:heuristics} we show how the QR representation of the reshaped Lindblad dynamics also leads to a suggestive heuristic argument upholding the bound.
\end{itemize}

\section{Non-Markovian dynamics can violate the bound}
\label{sec:nM}

We now turn to show how the bound (\ref{bound}) yields an experimentally realizable criterion for Markovian dynamics \cite{BrPe2002}. Recall that the most general evolution of a quantum system may be realized as a reduction of the unitary evolution of the ``system + environment''

\begin{align}
    \Lambda(t,0)(\bm{\rho}) = {\rm Tr}_{\rm E}\Big( \operatorname{U}(t) \bm{\rho} \otimes \bm{\rho}_{\rm E} \operatorname{U}^\dagger(t) \Big) ,
\end{align}
where $\bm{\rho}_{\rm E}$ is a state of environment  at the initial time $t=0$, and $\operatorname{U}(t)$ is a unitary operator acting on ``system + environment'' degrees of freedom \cite{BrPe2002,Alicki-Lendi,RiHu2012}. A quantum evolution represented by a dynamical map $\Lambda(t,s)$ is called Markovian if for any $t > s$ the propagator $$\Lambda(t,s):=\Lambda(t,0) \Lambda^{-1}(s,0) ,$$ is completely positive and trace-preserving. 
For any sequence $t_n>t_{n-1}> \ldots t_1 > 0$ we can write
\begin{align}
    \Lambda(t_n,0) = \Lambda(t_n,t_{n-1}) \circ %\Lambda_{t_{n-1},t_{n-2}} 
     \ldots \circ \Lambda(t_2,t_1)\circ \Lambda(t_1,0) ,  \label{Markov}
\end{align}
where `$\circ$' denotes the composition of maps. In the Markovian case, each of the factors is completely positive. 
Now, the most general form of a quantum master equation always admits a representation in terms of a time-dependent generator $\mathfrak{L}_t$ such that for any `$t$' $\mathfrak{L}_t$ takes the canonical form (\ref{LGKS}) with time dependent canonical couplings $\gamma_\ell(t)$ and decoherence operators $\operatorname{L}_\ell(t)$ \cite{HaCrLiAn2014}.  For sufficiently small time increments $\mathrm{d}t=t_{i}-t_{i-1}$, we can approximate each of the factors in (\ref{Markov}) with the exponential of $\mathfrak{L}(t_{i-1})\,\mathrm{d}t$ (Trotter formula cf. e.g. \cite{HalB2015}). Hence, if the $\gamma_\ell(t)$'s are positive for all times, i.e. the evolution is Markovian, the rate bound (\ref{bound}) holds true for each of the factors in (\ref{Markov}).  We may define local relaxation rates $\Gamma_\ell(t)$ as minus the real part of eigenvalues of $\mathfrak{L}(t)$. Note that knowing the dynamical map $\Lambda_{t,0}$ one may derive the structure of  the generator via 
 
 $$ \mathfrak{L}(t) = \dot{\Lambda}(t,0) \circ \Lambda^{-1}(t,0) . $$
By continuity of the eigenvalues, we generically infer that in the Markovian case
\begin{align}
    \Gamma_{d^2-1}(t) \leq \frac 1 d \sum_{n=1}^{d^2-1} \Gamma_n(t) . \label{bound-t}
\end{align}
In general, however, a completely positive dynamical map satisfies a master equation of the form (\ref{LGKS}) but for which some of the
$\gamma_\ell(t)$ can be (at least) temporally negative \cite{HaCrLiAn2014}. In such a case, the dynamical map is not completely positive divisible i.e. some or all of the factors in (\ref{Markov}) fail to be completely positive.
As a result the bound (\ref{bound-t}) can be (at least temporally) violated. Hence, (\ref{bound-t}) may be considered as an analog of the well known Leggett-Garg inequalities \cite{LG} which provide a necessary condition for ``classicality'', that is, whenever the measurement statistics can be explained in terms of a purely classical process, then Leggett-Garg inequalities are always satisfied. Hence, violation of Leggett-Garg inequalities (sometimes called temporal Bell inequalities) immediately implies that the considered process is genuinely quantum. Similarly, violation of (\ref{bound-t}) implies that the process is non-Markovian (meaning that there exists a pair $t > s$ for which the map $\Lambda_{t,s}$ is not completely positive).

\subsection{Example} 

To illustrate how the bound (\ref{bound-t}) can be used as a non-Markovianity witness let us consider the following  qubit evolution governed by (\ref{QUBIT-1}) with time dependent parameters $\omega(t)$ and $\gamma_i(t)$:

\begin{equation}\label{QUBIT}
  \mathfrak{L}(t)(\bm{\rho}) = - \imath\,\frac{\omega(t)}{2}[\sigma_z,\bm{\rho}] +\sum_{i=+,-,z}\gamma_i(t)\mathfrak{D}[\sigma_{i}](\bm{\rho}) .
\end{equation}
The non-vanishing local relaxation rates are
\begin{align}%\label{QUBIT:rates}
&\Gamma_{\rm L}(t) =  \gamma_+(t) + \gamma_-(t),  
\nonumber  \\
 & \Gamma_{\rm T}(t) = \frac{\gamma_+(t) + \gamma_-(t)}{2} + \gamma_z(t).   
\nonumber
\end{align}
The total rate equals to $\Gamma_{\rm L}(t) + 2 \,\Gamma_{\rm T}(t) $ and hence (\ref{bound-t}) reduces the following well known condition $2 \Gamma_{\rm T}(t) \geq \Gamma_{\rm L}(t)$, or, equivalently, in terms of local relaxation times

\begin{equation}\label{TT}
% 2\, T_{{\rm L}}(t) \geq  T_{\rm T}(t) .
 2\, T_{\rm L}(t) \geq  T_{\rm T}(t) ,
\end{equation}
{which is a time dependent version of (\ref{TT-1}).}
Hence, whenever (\ref{TT}) is violated the corresponding qubit evolution is non-Markovian. In the most extreme case
the bound (\ref{TT}) can be violated for all $t > 0$. Consider so-called eternally non-Markovian evolution \cite{HaCrLiAn2014} (cf. also \cite{Nina}) corresponding to

$$  \gamma_{\pm}(t) = 1 \ , \ \ \ \gamma_{z}(t) = - \frac 12 \tanh t .$$
Although one of the rates is negative, one finds \cite{HaCrLiAn2014,Nina} that the corresponding dynamical map $\Lambda_{t,0}$ is completely positive. In this case $\Gamma_{\rm L}(t) = 2$ and $\Gamma_{\rm T}(t) = 1 - \tanh t$, and hence $2\, T_{\rm L}(t) < T_{\rm T}(t)$ for all $t > 0$, that is, the bound (\ref{TT}) is {\em eternally} violated causing  the corresponding evolution to be {\em eternally} non-Markovian.

\section{Discussion and outlook}
\label{sec:outlook}

Before we conclude let us still make some additional comments. 

 \subsection{Compatibility with extensions}

 Let us observe that the bound satisfies the following self-consistency condition: a generator $\mathfrak{L}$ operates on the system's operators living in $\mathcal{B}(\mathcal{H})$. Suppose now that $\mathcal{H}$ is a linear subspace of a larger Hilbert space $\widetilde{\mathcal{H}}= \mathcal{H}\oplus \mathcal{H}_{\rm ext}$. Since $\mathfrak{L}$ can be represented via

\begin{align}
    \mathfrak{L}(\bm{\rho}) = \Phi(\bm{\rho}) + \operatorname{K}\bm{\rho} + \bm{\rho} \operatorname{K}^\dagger ,
\end{align}
with a completely positive map $\Phi(\bm{\rho}) = \sum_{\ell=1}^{d^{2}-1}\gamma_{\ell} \operatorname{L}_{\ell}\bm{\rho} \operatorname{L}^\dagger_\ell$, and $\operatorname{K}= -\imath \operatorname{H} - \frac 12 \sum_{\ell=1}^{d^{2}-1}\gamma_{\ell} \operatorname{L}^\dagger_{\ell} \operatorname{L}_\ell$, one may in a natural way extend $\mathfrak{L}$ to $\widetilde{\mathfrak{L}}$ acting on $\mathcal{B}(\widetilde{\mathcal{H}})$. Indeed, for any operator

$$ \widetilde{X} = \left( \begin{array}{cc} X & A \\ A^\dagger & B \end{array} \right) \, \in\, \mathcal{B}(\widetilde{\mathcal{H}}) ,$$
with $X : \mathcal{H} \to \mathcal{H}$, $A: \mathcal{H}_{\rm ext} \to \mathcal{H}$, and $B : \mathcal{H}_{\rm ext} \to \mathcal{H}_{\rm ext}$, one has

\begin{align}
    \widetilde{\mathfrak{L}}(\widetilde{X}) = \left( \begin{array}{cc} \mathfrak{L}(X) & \operatorname{K}A  \\  A^\dagger \operatorname{K}^\dagger & 0 \end{array} \right) .
\end{align}
$\widetilde{\mathfrak{L}}$ is a completely positive generator {\em artificially} extended to act on operators from  $\mathcal{B}(\widetilde{\mathcal{H}})$. Hence it possesses $D^2-1$ relaxation rates $\widetilde{\Gamma}_\alpha$ ($D=d+d_{\rm ext}$). Now comes the compatibility condition: if the relaxation rates of $\mathfrak{L}$ satisfy the bound (\ref{bound}), then relaxation rates of $\widetilde{\mathfrak{L}}$ satisfy

\begin{align}
    \widetilde{\Gamma}_\mu \leq \frac{1}{D} \sum_{\alpha=1}^{D^2-1} \widetilde{\Gamma}_\alpha , \ \ \ \ \mu=1,2,\ldots,D^2-1 .   \label{COMP}
\end{align}
We provide a simple proof of this statement in Appendix~\ref{app:ext}.

\subsection{Relation with the theory of Hurwitz polynomials}
\label{]sec:Hurwitz}

In linear stability analysis real polynomials only admitting roots with negative real part are called Hurwitz stable polynomials (see e.g. \cite{CopW1965} or \cite{BarY2008}). The eigenvalues of a Gorini-Kossakowski-Lindblad-Sudarshan generator are real or appear in complex conjugate pairs while always satisfying the conditions  (\ref{rates}). They are therefore the roots of a univariate Hurwitz stable polynomial. Canonical rates and decoherence operators determine the coefficients of the polynomial. It is therefore tempting to put forward a topological argument upholding the conjecture (\ref{bound}) by considering which kind of bifurcations may occur when continuously varying the coefficients of a Hurwitz polynomial.  Let the variation tread a path starting from a  case when the eigenvalue, say $ \lambda_{d^{2}-1}$, with most negative real part is actually real and non defective. In such a case the corresponding eigenoperator is self-adjoint. A straightforward application of Gershgorin's circle theorem \cite{VarR2004} allows us to prove that the bound (\ref{bound}) holds true in such a case.  Generically, we expect that only two kind of bifurcations occur while preserving Hurwitz stability. Either a real eigenvalue vanishes or a pair of real eigenvalues coalesce and then give rise to a pair of complex conjugate eigenvalues.  Only the latter bifurcation may affect the validity of (\ref{bound}). Let us suppose that $ \lambda_{d^{2}-1}$ undergoes such bifurcation. By considering the factorized form of the full characteristic polynomial, we may imagine that the bifurcation involving $ \lambda_{d^{2}-1}$ is effectively induced by a coefficient of order zero of a second order polynomial turning more negative than a critical threshold. In such a case, the value of the real part of $ \lambda_{d^{2}-1}$ after the bifurcation admits for lower bound the value of  $ \lambda_{d^{2}-1}$ before the bifurcation. Hence, the inequality (\ref{bound}) cannot be affected. 
This reasoning hints at the possibility of a direct algebraic proof of the conjecture also in the generic case when  $ \lambda_{d^{2}-1}$ is part of  a complex conjugate pair. If so, it also suggests the possible presence of further algebraic properties 
of the matrix expressing the reshaped form of the generator (see section~\ref{sec:Lyapunov}) that may deserve to be studied (see, e.g. \cite{GoSi1999}).

\subsection{Extension to time non-autonomous cases}
\label{sec:Lyapunov}

In section~\ref{sec:proof} we arrive at a tight bound on the relaxation rates by means of the Teich-Mahler construction of the Pauli master equation \cite{TeMa1992}. The construction only relies on self-adjoint preservation and therefore also holds  for time non-autonomous quantum master equations. 
This observation paves the way to the identification of a test or {\textquotedblleft}witness{\textquotedblright} of complete positivity
also for master equations amenable to the canonical form (\ref{LGKS})
but eventually with explicitly time dependent canonical couplings $\gamma_{\ell}(t)$ and decoherence operators $\operatorname{L}_{\ell}(t)$, with $\ell$ counting from $1$ to $d^{2}-1$ \cite{HaCrLiAn2014}. We then distinguish two cases.
\begin{itemize}
    \item The completely positive-divisible (Markovian) case defined by canonical couplings of positive functions of time:
 \begin{align}
 &	\gamma_{\ell}(t)\,\geq\,0 && \forall\,t\,\geq\,0
 \label{Lyapunov:positive}
 \end{align}
 and any $\ell=1,\dots,d^{2}$ (cf. Section \ref{sec:nM}),  
 \item  The non completely positive-divisible (non-Markovian) case when some of the $\gamma_{\ell}$'s are temporally negative.
\end{itemize} 
We also assume that in both cases canonical rates are continuous almost everywhere and bounded 
 \begin{align}
 	|\gamma_{\ell}(t)|\,<\,\infty && \forall\,t\,\geq\,0
 	\label{Lyapunov:bound}
 \end{align}
 and any $\ell=1,\dots,d^{2}$. The hypotheses ensure global existence and uniqueness of solutions which can be then regarded as the image of a completely bounded non-homgeneous semigroup (see, e.g. \cite{DoMG2023}). When (\ref{Lyapunov:positive}) also holds, the semigroup is in addition completely positive. 
As we wish to restrict our analysis to positive times, we find expedient to consider the auxiliary equation which we obtain by only changing the sign in front of the generator
 \begin{align}
& 	\dot{\bm{\rho}}(t)=-\mathfrak{L}(t)(\bm{\rho}(t)) && \forall\,t\,\geq\,0
 	\label{Lyapunov:aux}
 \end{align}
 In cases of physical interest we expect the dynamics of the auxiliary equation to be dominated by a positive maximal Lyapunov exponent $\chi_{d^{2}-1}$ for large values of the time variable $t$. In particular, when (\ref{Lyapunov:positive}) holds true, the dynamics is globally expansive. Self-adjoint preservation remains, in any case, intact.
 We can most transparently apply dynamical systems methods to our analysis if we reshape the operator dynamics \cite{MisJ2011,BeZy2006} and work with vectors in $\mathbb{C}^{d^{2}}$ by setting 
 \begin{align}
 %	\operatorname{res}(\bm{\rho}(t)) 
 |\bm{\rho}(t)\>\!\> = \sum_{i=1}^{d} \wp_{i}(t) \bm{\psi}_{i}(t)\otimes\bm{\bar{\psi}}_{i}(t). 
 	\nonumber
 \end{align}
Here, the $ \big{\{}\bm{\psi}_{i}(t)\big{\}}_{i=1}^{d}$ form an orthonormal basis of $\mathbb{C}^{d}$ at time $t$. After reshaping, (\ref{Lyapunov:aux}) becomes
 \begin{align}
 	|\dot{\bm{\rho}}(t)\>\!\> =-\mathscr{L}(t)|\bm{\rho}(t)\>\!\> ,
 	\nonumber
 \end{align}
 where $\mathscr{L}(t)$ denotes the reshaped generator $\mathfrak{L}(t)$. Evolution from initial conditions assigned at time zero is governed by a one parameter family of matrices $\mathscr{G}(t)$ in $\mathcal{M}_{d^{2}}(\mathbb{C})$ satisfying
 \begin{align}
 	& \dot{\mathscr{G}}(t)=-\mathscr{L}(t)\mathscr{G}(t) ,
\nonumber\\
& \mathscr{G}(0)=\mathrm{Id}_{d^{2}} .
 	\nonumber
 \end{align}
 A straightforward application of the
 inner product in $\mathbb{C}^{d^{2}}$ yields the identity between the Euclidean norms 
 \begin{align}
 	\left\| |\bm{\rho}(t) \>\!\> \right\|_{2} =\|\bm{\wp}(t)\|_{2} ,
 	\nonumber
 \end{align}
 where $\bm{\wp}(t)$ denotes the one parameter family of vectors with $d$-real components $\wp_{i}(t)$, as in section~\ref{sec:proof}, but now satisfies
 \begin{align}
 	\bm{\dot{\wp}}(t)=-\operatorname{W}(t)\bm{\wp}(t) .
 	\nonumber
 \end{align}
 The equality between the Euclidean norms implies
 \begin{align}
 	\begin{split}
	\chi_{d^{2}-1}&:=\lim_{t\uparrow\infty}\sup_{t}\frac{1}{t}\ln \left\| |\bm{\rho}(t) \>\!\> \right\|_{2}
 	\\
& 	=\lim_{t\uparrow
 		\infty}\sup_{t}\frac{1}{t}\ln\|\bm{\wp}(t)\|_{2}		. 		
 		\end{split}
 \label{Lyapunov:Lyapunov}
 \end{align}
 This means that we can extricate the largest Lyapunov exponent $\chi_{\max}$ by considering the associated $\bm{\wp}$-dynamics. We derive (\ref{Lyapunov:Lyapunov}) using the Euclidean norm but we already know that Lyapunov exponents are affine invariant indicators of the dynamics \cite{ByViGrNe1966,OseV1968,AdrL1995}. Hence  (\ref{Lyapunov:Lyapunov}) must hold for any Euclidean equivalent norm. We are therefore in the position to follow the same steps as in section~\ref{sec:proof}.  If the canonical couplings are always positive, i.e. (\ref{Lyapunov:positive}) holds true, we generically arrive at
 \begin{align}
 	\chi_{d^{2}-1}\,=\,\lim_{t\uparrow \infty} \sup_{t}\frac{1}{t}\ln \|\bm{\wp}(t)\| \,\leq\, \sup_{t \,\geq\,0}\sum_{\ell=1}^{d^{2}-1}\gamma_{\ell}(t) .
 	\label{Lyapunov:ineq1}
 \end{align}
We wish to turn this inequality into a relation between the maximal Lyapunov exponents and the other elements $\chi_{i}$ of the Lyapunov spectrum of (\ref{Lyapunov:aux}). 
A well known result of the theory of Lyapunov exponents states that
\begin{align}
	\begin{split}
&\lim_{t\uparrow \infty}	\sup_{t}\frac{1}{t}\ln |\det \mathscr{G}(t)| 
\\
&=\lim_{t\uparrow \infty}	\sup_{t}\frac{1}{t}\big{(} -\operatorname{Tr}(\mathscr{L}(t))\big{)} \,\leq\,
 \sum_{\ell=1}^{d^{2}-1}\lambda_{\ell} .
\end{split}
\nonumber
\end{align}
In writing the first equality we use the fact that the trace of the reshaped generator is real.  The inequality is tight for Lyapunov regular systems \cite{OseV1968,BeGaGiSt1980,DiRuVaVl1997}. 
The sum over the Lyapunov exponents starts from one because the Lyapunov spectrum inherits the property that at least one exponent vanishes in consequence of trace preservation.
We observe that
\begin{align}
	\sum_{\ell=1}^{d^{2}-1}\gamma_{\ell}(t)= - \frac{1}{d}\operatorname{Tr}(\mathscr{L}(t))
	\label{Lyapunov:trace}
\end{align}
holds true always. We conclude that when (\ref{Lyapunov:positive}) is verified, the Lyapunov exponents must satisfy
\begin{align}
	\chi_{d^{2}-1}\,\leq\,\frac{1}{d}\sum_{\ell=1}^{d^{2}-1}\chi_{\ell} .
	\label{Lyapunov:cp}
\end{align}
If we can only assume (\ref{Lyapunov:bound}) for all times then the upper bound (\ref{Lyapunov:ineq1}) on the maximal exponent takes the form
\begin{align}
	\chi_{d^{2}-1} \,\leq\, \sup_{t\,\geq\,0} \sum_{\ell=1}^{d^{2}-1}\,|\gamma_{\ell}(t)| .
	\nonumber
\end{align}
Using again (\ref{Lyapunov:trace}) we arrive at
\begin{align}
	\chi_{d^{2}-1} \,\leq\,\frac{\sum_{\ell =1}^{d^{2}-1}\chi_{\ell}}{d}+ \sup_{t\,\geq\,0}\sum_{\ell=1}^{d^{2}-1}\frac{|\gamma_{\ell}(t)|-\gamma_{\ell}(t)}{d} .
	\label{Lyapunov:cb}
\end{align}
The Lyapunov exponents $\chi_{i} $ are directly related to the real part of the logarithm of eigenvalues $\lambda_{i} $ of the flow solution of a linear system when the latter is time autonomous or periodic. The proof of the relation is straightforward in the former case whereas is a consequence of Floquet theory \cite{Pikovsky2016} in the latter case. 
The existence of a direct relation in more general cases has been only conjectured based on heuristic arguments \cite{GoKoSu1976}. 

The upshot is that for the time non-autonomous master equations, Lyapunov exponents become {\textquotedblleft}witnesses{\textquotedblright} of completely positive divisibility. Namely, (\ref{Lyapunov:cb}) showcases that evolution due to a non-completely positive semi-group may violate (\ref{Lyapunov:cp}). Violations generically occur when
\begin{align}
\sup_{t\,\geq\,0}\sum_{\ell=1}^{d^{2}-1}\frac{|\gamma_{\ell}(t)|-\gamma_{\ell}(t)}{d}\,>\,0 .
	\nonumber
\end{align}
Finally, we emphasize that whilst the supremum operation is generically needed in the definition of Lyapunov exponents, in may cases it can be replaced by a time average. In such cases violations of
complete positive divisibility corresponds to the condition
\begin{align}
	\lim_{t\uparrow \infty}\frac{1}{t}\int_{0}^{t}\mathrm{d}s\,\sum_{\ell =1}^{d^{2}-1}\frac{|\gamma_{\ell}(s)|-\gamma_{\ell}(s)}{d}\,>\,0 .
	\nonumber
\end{align}
The integrand coincides with the trace-norm rate of change introduced in \cite{RiHuPl2010} with the purpose of discriminating between a completely positive {divisible} (Markovian) from a non-completely positive {divisible} (i.e., non-Markovian) evolution. It is worth noticing that trace-norm rate of change
coincides with the Dahlquist-Lozinskii logarithmic ``norm'' \cite{DahG1958,LozS1958} (see Appendix~\ref{app:DL}) applied to the reshuffling or Choi matrix \cite{BeZy2006} of the reshaped dynamical map over an infinitesimal time interval.

{
\subsection{Open problems}
We stress that the bound (\ref{bound}) holds true for all GKLS master equations with finite dimensional Hilbert space. It 
may seem that in the limit $d \to \infty$ the r.h.s. of (\ref{bound}) vanishes as  $\frac 1d \to 0$. Note, however, that the number of relaxation rates $\Gamma_\ell$ increases as $d^2 \to \infty$ and hence still one may expect nontrivial bound which governs the relations between infinitely many rates. This problem definitely deserves further analysis. 
Another related question is what happens to the bound (\ref{bound}) if we relax the requirement of complete positivity? This issue might be interesting both from mathematical and physical point of view. In principle, one may consider a semigroup $\Lambda(t)$ consisting of positive trace-preserving maps or more generally $k$-positive maps where $k=1,\ldots,d$ (i.e. $k=1$ corresponds to semigroup of positive maps and $k=d$ corresponds to a semigroup of completely positive maps).
Now, $\mathfrak{L}$ generates a semigroup of $k$-positive maps if (cf. \cite{ChrD2022} for details)
\begin{equation}  \label{k-L}
   \< \bm{\psi}| ({\rm id}_k \otimes \mathfrak{L})(|\bm{\phi}\>\<\bm{\phi}|)|\bm{\psi}\> \geq 0 ,
\end{equation}
for all mutually orthogonal vectors $\bm{\psi},\bm{\phi} \in \mathbb{C}^k \otimes \mathcal{H}$. Here ${\rm id}_k$ denotes an identity map acting on the space of $k \times k$ complex matrices. What we have proved in this paper is that for $k=d$ the very condition (\ref{k-L}) implies the bound (\ref{bound}). It is natural to expect that for $k < d$  condition (\ref{k-L})  also implies some constraint for relaxation rates $\Gamma_\ell$. 
This bound might be also interesting in characterization of non-Markovianity. Namely, one calls \cite{DC+Sabrina} the dynamical map $\Lambda(t)$ $k$-divisible if all propagators $\Lambda(t,s)$ define $k$-positive maps for $t > s$. Hence, the violation of such bound immediately will imply that the evolution cannot be $k$-divisible.
}

    \section{Conclusions}
    \label{sec:finale}

    In this paper, we prove that the largest relaxation rate of the time autonomous Gorini-Kossakowski-Lindblad-Sudarshan master equation obeys the universal bound (\ref{bound}) as conjectured in \cite{ChKiKoSh2021}.
    As \cite{ChKiKoSh2021} already provides an example of Lindblad generator that saturates the bound, our result is optimal, that is, we show that  `$c=\frac 1d $' is the minimal constant for which the following constraint $\Gamma_{d^2-1} \leq c \sum_\alpha \Gamma_\alpha$ is true for all Lindblad generators in $d$-dimensions. This result,  therefore, accomplishes a long-term programme initiated in \cite{KimG2002}.

    The main tools we use in the proof are Teich-Mahler's derivation of the Pauli rate master equation \cite{TeMa1992} and elementary notions in the theory of Lyapunov exponents \cite{AdrL1995,CeCeVu2009}. Each particular solution $\bm{\rho}(t)$ of a quantum master equation specifies a distinct master rate equation for time-dependent populations (eigenvalues of $\bm{\rho}(t)$). For this reason, the Teich-Mahler-Pauli rate equation is the counterpart for quantum master equations of the mass conservation equation satisfied by particular solutions of a Fokker-Planck equation  written in terms of the current velocity \cite{NelE2001}. Indeed, as the current velocity plays a pivotal role in turning the second law of  classical (stochastic) thermodynamics  into an equality \cite{AuGaMeMoMG2012}, the Teich-Mahler-Pauli rate equation paves the way to derive fluctuation theorems for open master equations \cite{EsMu2006}. Our result constitutes a further illustration of the use of the Teich-Mahler-Pauli rate
    %Teich-Mahler
    equation as tool of theoretical analysis.

    The bound (\ref{bound}) is a relation between relaxation rates. These are quantities that can be directly measured in experiments. Hence, a universal bound on the largest relaxation rate paves the way to new operational tests of complete positivity complementary to that of \cite{WoEiCuCi2008} and with wide potential applications in quantum information processing. For instance, the recent paper \cite{Amato2024} uses the bound (\ref{bound}) to estimate the number of stationary states of completely positive semigroups.

To further underline the general relevance of our result we also consider the most general quantum evolution which goes beyond time autonomous Gorini-Kossakowski-Lindblad-Sudarshan master equation. In the general (non autonomous) scenario the relaxation rates $\Gamma_\ell(t)$ are time-dependent. Now, contrary to the autonomous case (Gorini-Kossakowski-Lindblad-Sudarshan semigroup) the time-dependent bound (\ref{bound-t}) may be violated (at least temporally). Interestingly, the violation of (\ref{bound-t}) has a clear physical meaning --- quantum evolution which violates (\ref{bound-t}) is non-Markovian meaning that it cannot represented as a composition of completely positive propagators.
Owing to its universality, the temporal bound (\ref{bound-t})  plays an analogous role to Bell inequalities and Leggett-Garg inequalities \cite{LG} (sometimes called temporal Bell inequalities). The violation of a Bell inequality rules out local hidden variable models, and the violation of Leggett-Garg inequalities rules out classical descriptions. Similarly, the violation of the relaxation bound rules out a completely positive Markovian dynamics.

\section{Acknowledgements} PMG is grateful to Antti Kupiainen for very insightful comments. DC was supported by the Polish National Science Center project No. 2018/30/A/ST2/00837. GK is supported by JSPS KAKENHI Grant Number 24K06873.

\appendix

\section{Compatibility with extensions}
\label{app:ext}

For simplicity of the analysis let us assume that we extend by a 1-dimensional subspace, i.e. we add an additional energy level. The proof for  a general (finite dimensional extensions)
goes along the same lines. If $\{|1\>,\ldots,|d\>\}$ defines an orthonormal basis in the original Hilbert space $\mathcal{H}$, then $\{|1\>,\ldots,|d\>,|d+1\>\}$ defines an orthonormal basis in $\mathcal{H} \oplus \mathcal{H}_{\rm ext}$, where $\mathcal{H}_{\rm ext}$ is spanned by $|d+1\>$.  Let

\begin{align}
	\operatorname{K}|\bm{e}_i \> = \kappa_i |\bm{e}_i\> \ , \ \ \ i=1,\ldots d,
\end{align}
where $\operatorname{K}= -\imath \operatorname{H} - \frac 12 \sum_{\ell=1}^{d^{2}-1}\gamma_{\ell} \operatorname{L}^\dagger_{\ell} \operatorname{L}_\ell$.
One finds for the spectrum of $\widetilde{\mathfrak{L}}$: for eigenvectors of the original $\mathfrak{L}$ one has

\begin{align}
	\widetilde{\mathfrak{L}}(X_\alpha) = \mathfrak{L}( X_\alpha) = \lambda_\alpha X_\alpha ,
\end{align}
and for $\widetilde{X}_i = |\bm{e}_i\>\<d+1|$

\begin{align}
	\widetilde{\mathfrak{L}}(\widetilde{X}_i) = \kappa_i \widetilde{X}_i \ , \ \  \widetilde{\mathfrak{L}}(\widetilde{X}_i^\dagger) = \overline{\kappa}_i \widetilde{X}_i^\dagger ,
\end{align}
together with $ \widetilde{\mathfrak{L}}(|d+1\>\<d+1|)=0 $. Hence, the relaxation rates $\widetilde{\Gamma}_\ell$ of $ \widetilde{\mathfrak{L}}$ read

$$  0,0=\Gamma_0,\Gamma_1,\ldots,\Gamma_{d^2-1},\frac 12 g_1,\ldots,\frac 12 g_d ,$$
where $g_i = - 2{\rm Re}\, \kappa_i$. Note that each $g_i$ is doubly degenerated. Hence, the sum of all rates equals

$$  \sum_{\alpha=1}^{(d+1)^2-1} \widetilde{\Gamma}_{\alpha}  = \sum_{\alpha=1}^{d^2-1} \Gamma_{\alpha} + \sum_{i=1}^d g_i . $$
Observe that $\frac 12 \sum_{i=1}^d g_i = - {\rm Tr} \operatorname{K} $. 
On the other hand
$$ {\rm Tr} \operatorname{K} = - \frac 12 \sum_{\ell=1}^{d^{2}-1}\gamma_{\ell} {\rm Tr}\Big(\operatorname{L}^\dagger_{\ell} \operatorname{L}_\ell\Big) = - \frac 12 \sum_{\ell=1}^{d^{2}-1}\gamma_{\ell} ,$$
and hence $\sum_{i=1}^d g_i = \sum_{\ell=1}^{d^{2}-1}\gamma_{\ell}$. Now, recalling
$ {\rm Tr}(- \mathfrak{L}) = d \sum_k \gamma_k$, one finds

\begin{equation}\label{}
	{\rm Tr}(- \widetilde{\mathfrak{L}}) =  {\rm Tr}(- \mathfrak{L}) + \sum_k g_k  = \left( 1 + \frac 1 d \right)  {\rm Tr}(- \mathfrak{L}) ,
\end{equation}
that is,

\begin{equation}\label{}
	\sum_{\ell=1}^{(d+1)^2-1} \widetilde{\Gamma}_\ell = \left( 1 + \frac 1 d \right)   \sum_{\ell=1}^{d^2-1} \Gamma_k ,
\end{equation}
which implies

\begin{equation}\label{}
	\frac{1}{d+1} \sum_{\ell=1}^{(d+1)^2-1} \widetilde{\Gamma}_\ell =  \frac 1 d   \sum_{\ell=1}^{d^2-1} \Gamma_k .
\end{equation}
Hence, if

\begin{equation}\label{}
	\frac 1d {\rm Tr}(- \mathfrak{L}) \geq  \Gamma_k ,
\end{equation}
then

\begin{equation}\label{}
	\frac{1}{d+1} {\rm Tr}(- \widetilde{\mathfrak{L}}) \geq  \widetilde{\Gamma}_k ,
\end{equation}
where $\widetilde{\Gamma}_k \in \{ \Gamma_1,\ldots,\Gamma_{d^2-1},g_1/2,\ldots,g_d/2\}$.

\section{Spectral properties of Lindblad generators}
\label{app:spectrum}

Any generator has the following spectral representation \cite{WolM2012,Kato1980,Facchi2019}

\begin{align}
    \mathfrak{L} = \sum_{\ell} ( \lambda_\ell \mathfrak{P}_\ell + \mathfrak{N}_\ell ) ,
\end{align}
where $\lambda_\ell$ are distinct eigenvalues of $\mathfrak{L}$,  $ \mathfrak{P}_\ell$ are projectors satisfying

$$   \mathfrak{P}_\ell  \mathfrak{P}_k = \delta_{k\ell}  \mathfrak{P}_\ell \ , \ \ \ \sum_\ell  \mathfrak{P}_\ell = {\rm Id} . $$
In terms of right $X_\ell$ and left $Y_\ell$ eigenvectors of $\mathfrak{L}$ one has $ \mathfrak{P}_\ell(X) = X_\ell {\rm Tr}(Y_\ell^\dagger X)$. It should be stressed that in general $ \mathfrak{P}_\ell$ are not self-adjoint, that is. $ \mathfrak{P}_\ell \neq  \mathfrak{P}_\ell^\ddagger$ (unless $\mathfrak{L}$ is normal, i.e. $ [\mathfrak{L},\mathfrak{L}^\ddagger]=0 $). Finally,   $\mathfrak{N}_\ell$ are nilpotent maps, i.e. $\mathfrak{N}_\ell^{n_\ell}=0$ for some integer $n_\ell$, satisfying

$$   \mathfrak{P}_\ell  \mathfrak{N}_k =  \mathfrak{N}_\ell  \mathfrak{P}_k =\delta_{k\ell}  \mathfrak{N}_\ell \ . $$
$\mathfrak{L}$ is diagonalizable (non defective) iff all $\mathfrak{N}_\ell=0$. The corresponding completely-positive semigroup has the following spectral representation

\begin{align}
    e^{t \mathfrak{L}} = \sum_\ell e^{t \lambda_\ell} \mathfrak{Q}_\ell(t) \mathfrak{P}_\ell ,
\end{align}
where

$$ \mathfrak{Q}_\ell(t) = e^{t \mathfrak{N}_\ell} = {\rm Id} + t \mathfrak{N}_\ell + \cdots + \frac{t^{n_\ell-1}}{(n_\ell-1)!} \mathfrak{N}_\ell^{n_\ell-1} ,  $$
due to $\mathfrak{N}_\ell^{n_\ell}=0$. Note, that $\mathfrak{Q}_\ell(t)$ are polynomial in `$t$' of degree $n_\ell-1$. One has therefore

\begin{align}
    e^{t \mathfrak{L}} = \sum_\ell \Big( e^{t \lambda_\ell}  \mathfrak{P}_\ell  + \sum_{k=1}^{n_\ell-1} \frac{t^{k}}{k!} \mathfrak{N}_\ell^{k} \Big) .
\end{align}
In the (generic) diagonalizable case it reduces to

\begin{align}
    e^{t \mathfrak{L}} = \sum_\ell e^{t \lambda_\ell}  \mathfrak{P}_\ell ,
\end{align}
that is, $e^{t \mathfrak{L}}(\bm{\rho}) = \sum_\ell e^{t \lambda_\ell} X_\ell {\rm Tr}(Y^\dagger_\ell \bm{\rho})$.

\section{The upper bound for the maximal Lyapunov exponent}
\label{app:bound}

To prove that  (\ref{X}) is indeed bounded from above by the unity for every $i=1,\dots,d$ and $n=1,\dots,d^{2}-1$, we now recall the orthonormality conditions (\ref{ortho}). They imply that there must exist a unitary
transformation relating the decoherence operators to the generators of $\mathfrak{su}(d)_{\mathbb{C}}\cong \mathfrak{sl}(d,\mathbb{C})$ \cite{HalB2015}.
This is because both collections of matrices are orthonormal bases of the space of traceless matrices on $\mathbb{C}^{d}$. According to this observation, we construct a basis of $\mathfrak{sl}(d,\mathbb{C}) $ adapted to the evaluation of the upper bound. We choose the first $d(d-1)$ as
\begin{align}
	&\mathfrak{g}_{\ell_{1}(i,j)}(t)=  |\bm{\psi}_{i}(t)\> \<\bm{\psi}_{j}(t) | && i\neq j
	\nonumber
\end{align}
with the convention that the lexicographic map $\ell_{1}(i,j) $ counts first outer products for $i\,<\,j$, ( i.e. $\ell_{1}(1,2)=1$ and  $\ell_{1}(d-1,d)=d\,(d-1)/2$ ) and then those for $i\,>\,j$
(i.e. $\ell_{1}(2,1)=d(d-1)/2+1$ and $\ell_{1}(d,d-1)=d(d-1)$). For completeness, we list the remaining basis elements
\begin{align}
	\mathfrak{g}_{\ell_{2}(i)}(t)= \frac{\sum_{j=1}^{i-1} |\bm{\psi}_{j}(t)\> \< \bm{\psi}_{j}(t)| -(i-1) |\bm{\psi}_{i}(t)\> \< \bm{\psi}_{i}(t)| }{\sqrt{i\,(i-1)}} ,
	\nonumber
\end{align}
where as $i$ varies between $2$ and $d$, the lexicographic map $\ell_{2}(i)  $ counts from $d\,(d-1)$ to $d^{2}-1$. This last set of basis elements consists of diagonal matrices that do not contribute to (\ref{X}). We arrive at the representation of the decoherence operators
\begin{align}
	\operatorname{L}_{n}=\sum_{i\neq j}^{d} \operatorname{U}_{n,\ell_{1}(i,j)}(t)\,\mathfrak{g}_{\ell_{1}(i,j)}(t) +\sum_{i=1}^{d-1}\operatorname{U}_{n,\ell_{2}(i)}(t)\,\mathfrak{g}_{\ell_{2}(i)}(t) .
	\label{unit}
\end{align}
Upon inserting (\ref{unit}) into (\ref{X}) we get
\begin{align}
	w_{n}^{(i)}(t) &=\sum_{l\neq i}^{d}\left(|\operatorname{U}_{n, \ell_{1}(l,i)}(t)|^{2}+|\operatorname{U}_{n, \ell_{1}(i,l)}(t)|^{2}\right) .
	\label{id}
\end{align}
The sum does not contain repetitions of the squared matrix elements of the unitary matrix $\operatorname{U}$. We conclude
\begin{align}
	w_{n}^{(i)}(t)\,\leq\,\sum_{\ell=1}^{d^{2}-1}|\operatorname{U}_{n,\ell}(t)|^{2}=1
	\nonumber
\end{align}
which implies
\begin{align}
	\chi=\Gamma_{d^{2}-1}\,\leq\,\sum_{\ell=1}^{d^{2}-1}\gamma_{\ell} .
	\nonumber
\end{align}

\subsection{Different norm can provide different bound.} 

Let us estimate the bound for $\chi$ using $\| \operatorname{W}(t) \|_1$. We find
\begin{align}
    \| \operatorname{W}(t) \|_1 = \max_j \sum_{i=1}^d |\operatorname{W}_{ij}(t) | = 2 \max_j  |\operatorname{W}_{jj}(t) | ,
\end{align}
due to $$\sum_{i=1}^d \operatorname{W}_{ij}(t) = 0$$ and $$\operatorname{W}_{ij}(t) \geq 0$$ for $i \neq j$. Furthermore we obtain
\begin{align}
     |\operatorname{W}_{jj}(t) | = \sum_{k \neq j} \operatorname{R}_{kj}(t) = \sum_n \gamma_n \sum_{k \neq j}  \left| \left\langle\,\bm{\psi}_{k}(t)\,,\operatorname{L}_{n}\bm{\psi}_{j}(t)\,\right\rangle%_{d}
     \right |^{2} , \nonumber
\end{align}
and using the same arguments as for the estimation of $\|\operatorname{W}(t)\|_\infty$, we arrive at
$$ |\operatorname{W}_{jj}(t) | \leq \sum_n \gamma_n ,$$
which results in

$$ \Gamma_{d^2-1} \leq \frac {2}{d} \sum_\ell \Gamma_\ell .$$
We thus recover the bound derived in \cite{WoCi2008} for purely dissipative generators.
This bound, however, contrary to (\ref{bound}) is not tight.

\section{Dahlquist-Lozinskii {\textquotedblleft}logarithmic norm{\textquotedblright}}
\label{app:DL}

The logarithm of the norm of the solution of a linear differential equation can be also estimated
by means of the Dahlquist-Lozinskii {\textquotedblleft}logarithmic norm{\textquotedblright} \cite{DahG1958,LozS1958}. Namely, for  any $\operatorname{A}=(a_{i,j})\in \mathcal{M}_{d}(\mathbb{C})$ and any matrix norm induced by an Euclidean equivalent vector norm the limit
\begin{align}
	\gamma(\operatorname{A}) :=\lim_{h \downarrow 0}	\frac{\left\| \operatorname{1}_{d}+h\,\operatorname{A}\right\|-1}{h}
	\label{DL:ln}
\end{align}
is well defined. The limit (\ref{DL:ln}) is commonly referred to as the logarithmic norm of  the matrix $\operatorname{A} $ although is not a positive definite quantity. It shares, however, with the norm the properties
\begin{align}
&	\gamma(c\,\operatorname{A})=c \,\gamma(\operatorname{A}) && \forall c\,\geq\,0
\nonumber\\
&\gamma(\operatorname{A}+\operatorname{B})\,\leq\,\gamma(\operatorname{A})+\gamma(\operatorname{B})
&& \forall \operatorname{B}\,\in \mathcal{M}_{d}(\mathbb{C})
	\nonumber
\end{align}
Furthermore, it is straightforward to verify  \cite{LozS1958} that the logarithmic norm gives an upper bound on the real part of the eigenvalues of $ \operatorname{A}$:
\begin{align}
&\operatorname{Re} \lambda	\,\leq\,\gamma(\operatorname{A}) && \forall\,\lambda\in \operatorname{Sp}\operatorname{A}
	\nonumber
\end{align}
We refer to \cite{BanJ2020} for a complete list of properties and corresponding derivation.
The connection between (\ref{DL:ln})  and the dynamics stems from the possibility to exchange the order of the $\lim$  and the  $ \sup$ operations
\begin{align}
&	\lim_{h \downarrow 0}	\frac{\sup_{\bm{x} | \|\bm{x}\|=1}\left\| (\operatorname{1}_{d}+h\,\operatorname{A})\bm{x}\right\|-1}{h}
	=
	\nonumber\\
&	\sup_{\bm{x} | \|\bm{x}\|=1}\lim_{h \downarrow 0}	\frac{\left\| (\operatorname{1}_{d}+h\,\operatorname{A})\bm{x}\right\|-\|\bm{x}\|}{h\,\|\bm{x}\|}. 
	\nonumber
\end{align}
See Appendix~1 of \cite{ByViGrNe1966} for a proof. The consequence is that
\begin{align}
	\frac{\mathrm{d}}{\mathrm{d} t^{+}}\ln \|\bm{\mathscr{x}}(t)\| \,\leq\,\gamma(\operatorname{A}(t))
	\nonumber
\end{align}
where left hand side must interpreted as the upper right Dini derivative of the logarithm of the norm (see, e.g. \cite{SodG2006,BanJ2020}). The inequality implies
\begin{align}
	\ln \|\bm{\mathscr{x}}(t)\|\,\leq\, \int_{0}^{t}\mathrm{d}s\,\gamma(\operatorname{A}(s))
	\nonumber
\end{align}
Similarly, from
\begin{align}
	-\gamma(-\operatorname{A})=\lim_{h \downarrow 0}	\frac{1-\left\| \mathsf{Id}-h\,\operatorname{A}(t)\right\|}{h}
\nonumber
\end{align}
we arrive at
\begin{align}
	\frac{\mathrm{d}}{\mathrm{d} t^{-}}\ln \|\bm{\mathscr{x}}(t)\| \,\geq\,-\gamma(-\operatorname{A}(t))
	\nonumber
\end{align}
and therefore
\begin{align}
	-\int_{0}^{t}\mathrm{d}s\,\gamma(-\operatorname{A}(s))	\,\leq\,\ln \|\bm{\mathscr{x}}(t)\|\,\leq\, \int_{0}^{t}\mathrm{d}s\,\gamma(\operatorname{A}(s)). 
	\nonumber
\end{align}
The logarithmic norm of a matrix can be straightforwardly calculated for the three most common norms \cite{ByViGrNe1966}. Namely using in (\ref{DL:ln}) the $\|\operatorname{A}\|_{1}$ matrix-norm yields
	$$ \gamma _{1}(\operatorname{A})=\sup \limits _{j}\left(\operatorname{Re} a_{j,j}+\sum \limits _{i\neq j}|a_{i,j}|\right)$$
If we instead use the spectral norm $\|\operatorname{A}\|_{2}$, we find
	$$ \gamma _{2}(\operatorname{A})=\max \left\{ \lambda \big{|} \lambda \in \operatorname{Sp}\left({\frac {\operatorname{A}+\operatorname{A}^{\mathrm {\dagger} }}{2}}\right)\right\}$$
whereas $\left\|\operatorname{A}\right\|_{\infty}$ yields
	$$ \gamma _{\infty }(\operatorname{A})=\sup \limits _{i}\left(\operatorname{Re} a_{i,i}+\sum \limits _{j\neq i}|a_{i,j}|\right)$$
It is straightforward to verify that our main result also stems from the application of this latter inequality to the Pauli master equation (\ref{TM}).

The original motivation of \cite{DahG1958,LozS1958} that led Dahlquist and Lozinskii to independently introduce the logarithmic norm of a matrix was the study of error bounds in numerical integration of differential equations. The concept has been later extended from matrix to bounded linear operators, and more recently to nonlinear maps and unbounded operators. We refer to \cite{SodG2006}
for an overview of recent developments.

\section{Heuristic derivation of the bound from QR representation and reshaped dynamics}
\label{app:heuristics}

 Our starting point is the reshaped Gorini-Kossakowski-Lindblad-Sudarshan generator \cite{Watrous2018,MisJ2011}
     \begin{align}
     	\begin{split}
     		&	\mathscr{L}=-\imath\operatorname{H}\,\otimes\,\operatorname{1}_{d}+\imath\,\operatorname{1}_{d}\otimes\,\operatorname{H}^{\top}+\sum_{n=1}^{d^{2}-1}\gamma_{n} \mathscr{D}_{n}
     		\\
     		&	\mathscr{D}_{n}=  \operatorname{L}_{n}\,\otimes\,\bar{\operatorname{L}}_{n}
     		-\frac{\operatorname{L}_{n}^{\dagger}\operatorname{L}_{n}\,\otimes\,\operatorname{1}_{d}+\operatorname{1}_{d}\,\otimes\,\operatorname{L}_{n}^{\top}\bar{\operatorname{L}}_{n}}{2}		
     	\end{split}
     	\label{L}
     \end{align}
     We are interested in the large negative time behavior so it is expedient to preliminarily redefine the time variable $t\mapsto -t$. Next we avail us of Theorem 2.1.14 of \cite{HoJo2013} and factorize the flow generated by (\ref{L}) into the product of an unitary $\mathscr{U}(t)$ and an upper triangular matrix $\mathscr{T}(t)$ with positive entries on the diagonal
     \begin{align}
     	e^{-\mathscr{L}t}=\mathscr{U}(t)\mathscr{T}(t)
     	\nonumber
     \end{align}
     This is the QR representation commonly applied in numerical algorithms for the computation of Lyapunov exponents \cite{BeGaGiSt1980, DiRuVaVl1997, SkoC2008,Pikovsky2016}.
     After standard manipulations, we arrive at
     \begin{align}
     	\mathscr{U}(t)^{\dagger}\dot{\mathscr{U}}(t)+\mathscr{\dot{T}}(t)\mathscr{T}(t)^{-1}=-\mathscr{U}(t)^{\dagger}\mathscr{L}\mathscr{U}(t)
     	\label{mat}
     \end{align}
     We observe that $ \mathscr{U}(t)^{\dagger}\dot{\mathscr{U}}(t)$ is equal to $\imath \,\mathscr{H}(t)$ where $\mathscr{H}(t)$ is a self adjoint matrix. Furthermore, the inverse of an upper triangular matrix, if it exists, is upper triangular and so is the product of upper triangular matrices.
     Hence diagonal entries of the matrix equation (\ref{mat}) yield
     \begin{align}
     	\ln \frac{\mathscr{T}_{i,i}(t)}{\mathscr{T}_{i,i}(0)}=-\int_{0}^{t}\mathrm{d}s\operatorname{Re}\sum_{l,k=1}^{d^{2}}\mathscr{\bar{U}}_{l,i}(s)\mathscr{L}_{l,k}\mathscr{U}_{k,i}(s)
     	\nonumber
     \end{align}
     We thus recover the equation commonly used for numerical evaluation of Lyapunov exponents via the so-called \textquotedblleft QR\textquotedblright\ factorization.
     If we now heuristically argue that time averaging over an infinite time horizon suppresses interference we arrive at
     \begin{align}
     &	\chi\,\leq\, \max_{1\,\leq\,\ell\,\leq\,d^{2}} \mathscr{L}_{\ell,\ell}
     \nonumber\\
&     	=\max_{1\,\leq\,i,j\,\leq\,d} \operatorname{Re}\left \langle\,\bm{f}_{i}\otimes\bm{\bar{f}}_{j}\,,(-\mathscr{L})\bm{f}_{i}\otimes\bm{\bar{f}}_{j}\,\right\rangle%_{d^{2}}
     	\nonumber
     \end{align}
     where the $\bm{f}_{i} $'s are the elements of an arbitrary orthonormal basis of $\mathbb{C}^{d}$.  After straightforward algebra we get
     \begin{align}
     &	\operatorname{Re}\left \langle\,\bm{f}_{i}\otimes\bm{\bar{f}}_{j}\,,(-\mathscr{L})\bm{f}_{i}\otimes\bm{\bar{f}}_{j}\,\right\rangle%_{d^{2}}
     =
     	\nonumber\\
&     \sum_{k=1}^{d^{2}-1}\gamma_{k} 	\begin{cases}
     		\sum_{r\neq i}^{d} \,|z_{r,i}^{(k)}|^{2} &  i=j
     		\\[0.3cm]
     		\frac{\sum_{r\neq i}^{d} \,|z_{r,i}^{(k)}|^{2}+\sum_{r\neq j}^{d}|z_{r,j}^{(k)}|^{2}  }{2}	+\frac{|z_{i,i}^{(k)} -\bar{z}_{j,j}^{(k)}|^{2}}{2}& i\neq j
     	\end{cases}
     	\nonumber
     \end{align}
     with
     \begin{align}
     	z_{r,i}^{(k)}=\left \langle\,\bm{f}_{r}\,,\operatorname{L}_{k}\bm{f}_{i}\,\right\rangle%_{d}
     	\nonumber
     \end{align}
     From this point, we recover the bound by estimating the sum over squared matrix elements as in the proof of $w^{(n)}_i(t) \leq 1$.

\vspace{.2cm}

	\bibliography{GKLS_rates}{} % the path must be given in reference to the compilation directory
	\bibliographystyle{apsrev4-2}
\end{document}